\documentclass[journal]{IEEEtran} 
\usepackage{hy} 
\usepackage[linesnumbered,lined, algoruled]{algorithm2e}
\usepackage{multirow}
\usepackage{tikz}
\usetikzlibrary{arrows}
\usepackage{hyperref}
\graphicspath{{./LO}}
\def\minim{\hbox{min }}
\def\wrt{\hbox{w.r.t. }}
\def\st{\hbox{s.t. }}

\renewcommand{\Re}{\mathop{\rm Re}\nolimits}

\ifbozze
                 
\fi

\begin{document}

\labelfigure{./}

\title{Cooperative Localization in WSNs:\\ a Hybrid Convex/non-Convex Solution}

\author{%
Nicola~Piovesan, Tomaso~Erseghe
\thanks{The authors are with the Dipartimento di Ingegneria dell'Informazione, Universit\`a di Padova, Via G. Gradenigo 6/B, 35131 Padova, Italy. Contact author: Tomaso Erseghe, tel: +39 049 827 7656, fax: +39 049 827 7699, mailto: erseghe@dei.unipd.it
} %
} 

\markboth{Padova, \today \hfill SUBMITTED TO IEEE TRANSACTIONS ON SIGNAL AND INFORMATION PROCESSING OVER NETWORKS\hfill }{}
\maketitle
\ifbozze\clearpage\fi


\acrodef{ADMM}{alternating direction method of multipliers} 
\acrodef{APP}{auxiliary problem principle} 
\acrodef{AWGN}{additive white Gaussian noise} 
\acrodef{CPLD}{constant positive linear dependence}
\acrodef{CRLB}{Cramer-Rao lower bound} 
\acrodef{DC}{direct current} 
\acrodef{DP}{dynamic programming} 
\acrodef{IPM}{interior point method} 
\acrodef{KKT}{Karush Kuhn Tucker} 
\acrodef{LMS}{least mean squares} 
\acrodef{MAP}{maximum a posteriori} 
\acrodef{MEMS}{micro electro-mechanical system} 
\acrodef{ML}{maximum likelihood} 
\acrodef{OPF}{optimal power flow} 
\acrodef{PCC}{point of common coupling} 
\acrodef{PDF}{probability density function} 
\acrodef{PEI}{power electronic interface} 
\acrodef{RMSE}{root mean squared error} 
\acrodef{RF}{radio frequency} 
\acrodef{RSSI}{received signal strength indicator} 
\acrodef{SDP}{semi-definite programming}
\acrodef{SNR}{signal to noise ratio} 
\acrodef{SMG}{smart micro grid} 
\acrodef{TOA}{time of arrival} 
\acrodef{TOF}{time of flight} 
\acrodef{WSN}{wireless sensor network}

\begin{abstract}
We propose an efficient solution to peer-to-peer localization in a wireless sensor network which works in two stages. At the first stage the optimization problem is relaxed into a convex problem, given in the form recently proposed by Soares, Xavier, and Gomes. The convex problem is efficiently solved in a distributed way by an ADMM approach, which provides a  significant improvement in speed with respect to the original solution. In the second stage, a soft transition to the original, non-convex, non relaxed formulation is applied in such a way to force the solution towards a local minimum. The algorithm is built in such a way to be fully distributed, and it is tested in meaningful situations, showing its effectiveness in localization accuracy and speed of convergence, as well as its inner robustness.
\end{abstract}


\section{Introduction}

\IEEEPARstart{S}{mart sensors}  nowadays have grown rapidly thanks to the proliferation of \ac{MEMS} technology and advances in \ac{RF} communications. This kind of sensors are the basic unit of a \ac{WSN}, which are extending our ``\emph{ability to monitor and control the physical world}.'' Sensor nodes can in fact ``\emph{sense, measure and gather information from the environment and, based on some local decision process, they can transmit the sensed data to the user}'' \cite{yick2008wireless}. They are characterized by their reduced dimension, limited processing and computing resources, and low costs. \acp{WSN} can be used either for monitoring applications (which include  indoor/outdoor environmental monitoring, seismic monitoring, health monitoring), but also for tracking applications (of objects, animals, humans and vehicles). In this latter context, cooperative localization in \acp{WSN} is a task that has gained increasing interest, especially in indoor scenarios where satellite communications cannot be employed, but also for the inner capability of providing more accurate results. The possibility of efficiently locating the objects opens up a wide number of applications ranging from the industrial to the health-care environment, and it is also beneficial for an efficient management of the communication network itself \cite{cadger2013survey}. However, it also calls for suitably fast and simple solutions.

In this scenario, the localization problem we aim at solving is one where a number, $N$, of nodes with ranging capability wirelessly exchanges ranging measurements with its neighboring nodes, i.e., those inside a given communication radius. By letting $\B p_i$ be the position of the $i$th node, $\C N_i$ the set of neighbor nodes to which it communicates and for which a ranging measurement is available, and $r_{i,j}$ the noisy ranging measurement  between node $i$ and its neighbor $j\in\C N_i$, then the optimization problem we wish to solve is of the form 
$$
\eqalign{
& \minim \sum_{i\in\C N,\,j\in\C N_i} \fract12 \big(\|\B p_i-\B p_j\| - r_{i,j} \big)^2 \cr
& \wrt \B p_i, i\in\C N=\{1,\ldots,N\}\cr
& \st \B p_k = \B a_k,\; k\in\C A\,.
}
\e{HY2}
$$
In the problem formulation we implicitly considered that the ranging measurement is corrupted by an \ac{AWGN}, and  we also considered that a subset $\C A\subset\C N$ of the nodes are \emph{anchor} nodes, i.e., nodes for which the exact position $\B a_i$ is known. In the intended scenario, only a very few  nodes are anchors (e.g., the all-but-one-node are anchors is not the intentional scope of our investigation), which makes the problem highly non-convex, and in general difficult to solve. 

The above problem has been previously considered, e.g., in \cite{Langendoen:cn03, Cheung:tsp05, Denis:tmtt06, Mao:cn07, Shi:tsp10, Wang:sjo08, Cota:sj13, Simonetto:tsp14, Soares:tsp15, Erseghe:sipn15}. We concentrate on the more relevant solutions. The approach considered in \cite{Wang:sjo08, Shi:tsp10, Simonetto:tsp14} uses a \ac{SDP} relaxation in order to map the non-convex original problem into a convex problem. Although the idea is potentially interesting, especially because a convergence guarantee can be obtained, the \ac{SDP} relaxation implies a non trivial computational effort which makes it impractical in large networks. A valid alternative is provided by the convexification method proposed by Soares, Xavier, and Gomez in \cite{Soares:tsp15}. By exploiting the concept of convex envelope, the authors are able to identify an algorithm which is simple and scalable. Since all of the above mentioned solutions are expected to provide an identical performance, \cite{Soares:tsp15} should be considered the preferred convex relaxation approach. Its main drawback, however, is the lack of adherence to the original non-convex problem, which practically means that the solutions are not guaranteed to be global, nor local, minima. In this interpretation, working directly on the non-convex problem \e{HY2} might be a viable option. This option was investigated by Erseghe in \cite{Erseghe:sipn15}, which proposes a solution based upon the \ac{ADMM} method, suitably modified in order to guarantee convergence to a local minimum also in the presence of harsh non-convexities. The plain \ac{ADMM} solution is in fact known to converge only under convex problem formulations. The resulting algorithm is effective and implies a controlled computational burden.  The main drawback with this latter method is, however, a general difficulty to ensure convergence to a good local minimum, especially in a worse case situation where the starting point is very far from optimum, and the problem is highly non-convex. 

In this paper, we aim at bringing together the positive aspects of \cite{Soares:tsp15} and \cite{Erseghe:sipn15}, by proposing an hybrid solution to the \ac{WSN} cooperative localization problem \e{HY2}. The leading idea is to exploit the convex relaxation method introduced in \cite{Soares:tsp15} in order to identify a starting solution which is then refined -- and guaranteed to be at least a local optimum -- by using the ideas developed in \cite{Erseghe:sipn15}. By using a relaxation we also expect to obtain a faster convergence rate than with the standard non-convex approach, the reason simply being the higher level of coordination ensured by removing non-convexities. The transition between the two approaches is meant to be smooth, and to be implementable in a fully distributed fashion. To this aim, the \ac{ADMM} method is used to solve both the original problem as well as its convex counterpart. More precisely, the application of \ac{ADMM} to the convex relaxation problem of \cite{Soares:tsp15} will be shown to provide a significant improvement over the parallel Nesterov's method originally proposed by the authors, that is, a much faster convergence rate. The activation of the non-convex function under local convergence of the relaxation will further ensure convergence to a local optimum. This gives a gain in localization accuracy over \cite{Soares:tsp15}, which will be shown to be rather significant. The localization accuracy of the proposed solution is equivalent to the one that can be obtained by \cite{Erseghe:sipn15}, but with two relevant enhancements: a faster convergence speed and, even more importantly, a strong resilience with respect to the parameters choice which improves the algorithm robustness.

The paper is organized as follows. The convex relaxation is formalized in \sect{CR}, the problem is put in a networked form which is suitable for being implemented in a distributed fashion in \sect{PF}, the overall distributed algorithm is presented in \sect{AL}, and its computational complexity is discussed in \sect{CC}. Performance is assessed in \sect{SI}, where a comparison with state-of-the-art solutions is also given. \sect{CO} concludes the paper.

\Section[CR]{Convex Relaxation}

As we discussed, the constituent functions of \e{HY2} are of the form
$$
F_i(\B x_i) =  \sum_{j\in\C N_i} \fract12 \big(\|\B p_i-\B p_j\| - r_{i,j} \big)^2\;,
\e{HY4}
$$
where $\B x_i = \{\B p_j\}_{j\in\{i\}\cup \C N_i}$ collects copies of the position of node $i$ together with the positions of its neighbors. Note that, unlike \cite{Soares:tsp15}, the approach we are using in \e{HY2} does not involve two different notations for anchor and non-anchor nodes, and therefore it compacts and simplifies exposition. 

Now, according to \cite{Soares:tsp15}, the contribution
$$
f(\B z, r) = \fract12 (\|\B z\| - r )^2
\e{HY10}
$$
can be equivalently written in the form
$$
f(\B z,r) = \min_{\Bs y:\;\|\Bs y\|=r} \fract12 \|\B z-\B y\|^2\;.
\e{HY12}
$$
This provides the convex envelope through a simple relaxation of the non-convex constraint $\|\B y\|=r$ into the convex constraint $\|\B y\|\le r$. To better evidence the convexity property, we can equivalently write $\|\B y\|^2\le r^2$. We therefore obtain a convex relaxation based upon contributions of the form
$$
\tilde{f}(\B z,r) =\min_{\Bs y:\;\|\Bs y\|\le r} \fract12 \|\B z-\B y\|^2\;,
\e{HY14}
$$
the convexified counterpart to \e{HY4} being
$$
\tilde{F}_i(\B x_i) =  \sum_{j\in\C N_i} \tilde{f}\big(\B p_i-\B p_j, r_{i,j}\big)\;.
\e{HY8}
$$

We observe that $\tilde{f}$ can be given in an explicit form, which reveals the simplicity of the convex relaxation, and which was not evidenced in \cite{Soares:tsp15}. Now, the function in \e{HY14} reaches its minimum when $\B y = \B z$. This solution is not always licit but we can say that the optimum solution corresponds to the closest allowed point $\B y$ to $\B z$. We can therefore distinguish between two cases:
\begin{enumerate}
\item If $\|\B z\|\le r$, then $\B z$ is a member of the set from which we can select $\B y$, and the value that minimizes the function is simply $\B y = \B z$. In this case, the function value is simply $\tilde{f}(\B z,r) =0$.
\item If, instead, $\|\B z\|> r$, then the value of $\B y$ that minimizes \e{HY14} is on the boundary of the circle $\|\B y\| = r$ and, more specifically, it corresponds to the intersection between the circle and the line that connects the point $\B z$ to the origin. We therefore have $\B y=r \B z/\|\B z\|$, and an outcome equal to $\tilde{f}(\B z,r) =\fract12(\|\B z\|-r)^2$.
\end{enumerate}
To summarize these results, we can write
$$
\tilde{f}(\B z,r) = g(\|\B z\|-r) \;,
\e{HY16}
$$
where
$$
g(x) = \fract12 x^2\cdot1(x)\;,
\e{HY18}
$$
and where $1(x)$ is the unit step function, providing $1$ for $x\ge0$ and $0$ for $x<0$. Note  a close relation between \e{HY12} and \e{HY16}, the latter simply setting to zero the result when $\|\B z\|<r$. An illustration of the effect given by the convex approximation can be found in \cite[Fig.~1]{Soares:tsp15}. Note also that $g$ has first and second derivatives 
$$
\eqalign{
g'(x)& =x\cdot1(x)=[x]^+\cr
g''(x)&=1(x)
}
\e{HY20}
$$
which will be useful later on in order to identify gradients and Hessians.

\Section[PF]{Problem Formalization}

In order to approach the solution of \e{HY2}, or of its convex counterpart replacing \e{HY4} with \e{HY8}, the decomposition and coordination method of \cite{Erseghe:sipn15} is employed. In this context, problem \e{HY2} is put in an equivalent form where variables $\B p_i$ are duplicated in such a way that the generic node $i$ owns its copy of variables $\B x_i=\{\B p_j\}_{j\in\C N_i\cup\{i\}}$. Specifically, problem \e{HY2} assumes the form
$$
\eqalign{
&\minim F(\B x)\cr
&\wrt \B x\in\C X, \B z\in\C Z\cr
& \st \B A\B x = \B z\,,
}
\e{AL2}
$$
where $\B x=[\B x_1,\ldots,\B x_N]$, with $\B x_i=[\B x_{i,j}]_{j\in\C N_i\cup\{i\}}$ collecting in its entries the replicas of the position of node $i$, namely, $\B x_{i,i}=\B p_i$, and those of its neighbors $\B x_{i,j}=\B p_j$, $j\in\C N_i$. If we denote with $n$ the coordinate dimension, namely, $n=2$ for 2D localization, and $n=3$ for 3D localization, then $\B x_i$ has length $n(1+N_i)$, with $N_i$ the cardinality of $\C N_i$, i.e., the number of neighbors of node $i$. The target function in \e{AL2} is the separable function
$$
F(\B x) = \sum_{i\in\C N} F_i^\bullet(\B x_i)\;,
\e{AL3}
$$
with $F_i^\bullet=F_i$ as defined in \e{HY4} if the original non-convex formulation is used, and $F_i^\bullet=\tilde{F}_i$ as defined by \e{HY8} if the convex relaxation is used. Moreover, set $\C X$ assumes the cartesian form $\C X=\C X_1\times\ldots\times\C X_N$, where each $\C X_i$ is itself separable in the form 
$$
\C X_i = \C R_i \times \Big\{ \bigotimes_{j\in \C N_i} \C R_j\Big\}\;,\quad
\C R_i = \cases{\C R& $i\not\in\C A$\cr \{\B a_i\}& $i\in\C A$\,,}
\e{AL3bis}
$$
with $\C R$ any (bounded) region containing the nodes positions.

The equivalence between replicas of the same position is jointly ensured in \e{AL2} by the constraint $\B A\B x=\B z$ and by the fact that $\B z\in\C Z$. Specifically, the linear constraint is locally given in the form\footnote{Note in the comparison with \cite{Erseghe:sipn15} that in \e{AL4}-\e{AL6} parameters $\epsilon$ and $\zeta$ have been dropped, i.e., they have been set to value $1$, since we verified that this simplification does not affects the final algorithm performance.}
$$
\B A_i\B x_i = \B z_i =\qmatrix{\B z_i^-\cr\B z_i^+}
\e{AL4}
$$
with
$$
\B A_i = \qmatrix{\B 1_{N_i} & -\B I_{N_i}\cr\B 1_{N_i} & \phantom{-}\B I_{N_i}\cr} \otimes\B I_n\;,
\e{AL6}
$$
where $\otimes$ is the Kronecker product, $\B 1_k$ denotes a column vector of length $k$ with all its entries set to $1$,  $\B I_k$ denotes the identity matrix of order $k$, and  $n$ is the coordinate dimension. For consistency, in the above we assumed that $\B z=[\B z_1,\ldots,\B z_N]$ is the collection of local contributions, and that $\B z_i^-=[\B z_{i,j}^-]_{j\in\C N_i}$ and $\B z_i^+=[\B z_{i,j}^+]_{j\in\C N_i}$. As a consequence, the length of both $\B z_i^-$ and $\B z_i^+$ is $nN_i$. It also is
$$
\B A = {\rm diag}(\B A_i,i\in\C N)\;.
\e{AL7}
$$
Note that \e{AL4} separately identifies differences of the form $\B z_{i,j}^-=\B p_i-\B p_j$, which are the ones effectively used in the target function \e{HY4}, and contributions of the form $\B z_{i,j}^+=\B p_i+\B p_j$. These are made consistent throughout the network by forcing $\B z$ to belong to the linear space
$$
\C Z = \Big\{\B z\Big|\B z_{i,j}^- = -\B z_{j,i}^-, \B z_{i,j}^+ = \B z_{j,i}^+, \forall i\in\C N,  j\in\C N_i\Big\}\;.
\e{AL8}
$$

The approach described so far is redundant in that it identifies $2nN_i$ constraints, $\B A_i\B x_i=\B z_i$, for $n(1+N_i)$ scalar variables, $\B x_i$. However, a number of reasons make it desirable. It is in fact particularly well suited for distributed implementation, and, more importantly, it provides an advantage in terms of convergence speed. The reason for the latter originates from the use of variables $\B z_{i,j}^-=\B p_i-\B p_j$ which allow to treat separately the convergence on relative positions $\B z_{i,j}^-$ (which set the target value), from the convergence with respect to absolute positions $\B z_{i,j}^+$ (which set the final localization outcome). Further details on this idea can be found by the interested reader in \cite{Erseghe:eurasip15,Erseghe:sipn15}.

\Section[AL]{Distributed Algorithm}

A distributed algorithm can be obtained by applying the \ac{ADMM} concept to \e{AL2}. The \ac{ADMM} is a simple but powerful algorithm that solves optimization problems decomposing them into smaller local sub-problems, which are easier to handle. The solutions to these local subproblems are coordinated in order to find the solution to a global problem. This algorithm is well suited for distributed optimization and in the latest years it found several applications in different areas \cite{Bertsekas:97, Boyd:10}.

The \ac{ADMM} method we use is taken from \cite{Erseghe:sipn15} and, in the present context, provides \alg{ALO4}.\Alg[h]{ALO4} Specifically, in \alg{ALO4} saddle points of the augmented Lagrangian 
$$
\eqalign{
L(\B x,\B z,\B\lambda,\B c) & = \sum_{i\in\C N} F_i^\bullet(\B x_i) + \langle\B\lambda_i,\B A_i\B x_i-\B z_i\rangle 
\cr & \hspace*{20mm} 
+ \fract12c_i\|\B A_i\B x_i-\B z_i\|^2
}
\e{AL20}
$$
are searched for by an alternate search that separately optimizes for each of the variables $\B x$, $\B z$, and $\B\lambda$, where:
\begin{enumerate}
\item The optimization with respect to $\B x$ is obtained in line~8.
\item The optimization with respect to $\B z$ is obtained in lines~10-13 through a local exchange of information. Note that messages $\B m_i$ have the same structure of $\B z_i$, that is $\B m_i=[\B m_i^-,\B m_i^+]$.
\item The update of $\B\lambda$ is obtained in line~15, where $\C P_{\lambda_{\max}}$ performs a clipping of vector entries in the range $[-\lambda_{\max};\lambda_{\max}]$, but any other clipping method can be used. 
\end{enumerate}
In line~19, if the non-convex formalization is used, the procedure updates the penalty parameters $\B c$ under two different conditions, namely: 1) if the primal gap $\B A_i\B x_i-\B z_i$ does not decrease sufficiently, where the chosen measure corresponds to an infinity norm criterion -- i.e., maximum value--; and 2) if the penalty parameters of neighbors have been previously increased. Note that the parameters used in \e{AL24} must satisfy $\delta_c>1$ and $0<\theta_c<1$, but a reliable algorithm is obtained only with $\delta_c\gtrsim 1$ and $\theta_c\lesssim1$. 

A smooth transition between the convex relaxation employing functions $\tilde{F}_i$, and the non-convex original formulation employing functions $F_i$, is simply managed by starting from a non-convex formulation on every node (line~3), and by locally activating the non-convex functions as soon as the local primal gap exceeds a given threshold (lines~21-22). If constant $\tau_c$ is chosen sufficiently small, then the non-convex formalization is activated only after convergence is reached on the convexified problem.

To summarize, the parameters used in \alg{ALO4} are:
\begin{enumerate}
\item The value $\epsilon_c$ for penalty parameters to be used with the convex formalization (line~6), and the initial value $\zeta_c$ for penalty parameters to be used with the non-convex formalization (line~23). These are the most relevant parameters that set the convergence speed and that must be wisely chosen according to the network characteristics. These values are kept separate since the non-convex functions $F_i$ imply larger function values (e.g., see the pictorial representation in later \fig{HY2}), hence larger values of $c_i$ to correctly balance function value and equality constraint in \e{AL16}. For this reason we must also set $\zeta_c\gg\epsilon_c$. As detailed in \cite{Erseghe:sipn15}, small variations in these values do not affect performance, provided that they are chosen in the correct range.
\item The clipping range $\lambda_{\max}$ for Lagrange multipliers (line~15). The parameter should be chosen sufficiently high, in order to prevent unwanted clipping actions. The standard choice, which will be used later on in the numerical simulations section, is $\lambda_{\max}=10^3$.
\item The update parameters $\delta_c$ and $\theta_c$ for penalty parameters (line~19). Standard choices, which will be used later on in the numerical simulations section, are $\delta_c=1.01$ and $\theta_c=0.98$.
\item The threshold $\tau_c$ for activating the original non-convex problem (line~21). This is another very relevant parameter that must be adequately chosen in dependence of the considered network in order to speed up convergence.
\end{enumerate}

As proved in \cite{Erseghe:sipn15}, which, in turn, derives from the ideas developed in \cite{Andreani:siam07, Martinez:siam14} about practical Lagrange methods, the fact that we bound both primal variables, $\B x$ and $\B z$, as well as Lagrange multipliers, $\B\lambda$, ensures that \alg{ALO4} will converge. If the considered functions $F_i$ were convex, then the limit point of \alg{ALO4} would identify a global minimum. In the non-convex scenario, however, the algorithm may find a local, rather than a global, minimum.

\Section[CC]{Computational Complexity Considerations and Further Insights}

We observe that \alg{ALO4} involves very simple operations, except for the local update \e{AL16} which corresponds to an optimization problem of order $n(1+N_i)$. The problem can be approached via standard optimization techniques relying on gradients and Hessians, which can be compactly expressed in the form
$$
\eqalign{
\nabla F_i^\bullet(\B x_i) & = 
 \qmatrix{\displaystyle \sum_{j\in\C N_i} \B A_{i,j}
\cr\rule{0mm}{4mm} \displaystyle-[\B A_{i,j}]_{j\in\C N_i}} \cr
\nabla^2 F_i^\bullet(\B x_i) & = 
 \qmatrix{\displaystyle \sum_{j\in\C N_i} \B B_{i,j} &  \displaystyle-[\B B_{i,j}]^T_{j\in\C N_i}
\cr\rule{0mm}{4mm} \displaystyle-[\B B_{i,j}]_{j\in\C N_i} &  \displaystyle{\rm diag}([\B B_{i,j}]^T_{j\in\C N_i})},\cr\,
}
\e{AL32}
$$
with 
$$
\eqalign{
\B A_{i,j} &= A(\B x_{i,i}-\B x_{i,j},r_{i,j})\cr
\B B_{i,j} &= B(\B x_{i,i}-\B x_{i,j},r_{i,j})\;,
}
\e{AL33}
$$
and where we used
$$
\eqalign{
 A(\B z,r) & = \frac{\B z}{\|\B z\|} [\|\B z\|- r ]^\bullet\cr
B(\B z,r) & = \frac{\B I_n }{\|\B z\|} [\|\B z\|- r ]^\bullet + \frac{\B z\B z^T}{\|\B z\|^3} r \cdot 1^\bullet(\|\B z\|- r )\;.
}
\e{AL34}
$$
with 
$$
[x]^\bullet = x\cdot 1^\bullet(x)\;,\qquad 1^\bullet(x) = \cases{
1 & if $F_i$\cr
1(x) & if $\tilde{F}_i$\cr
}
\e{AL62b}
$$
to take into account both the convexified as well as the non-convex case.

However, when region $\C R$ is sufficiently large and ranging measurements are sufficiently reliable that we can drop the constraint given by $\C R$, then the minimization problem  \e{AL16} entails a simplified version. This possibility is a powerful result that was not discussed in  \cite{Erseghe:sipn15}, and which we now separately address in case node $i$ \emph{is an anchor} and in case node $i$ \emph{is not an anchor}. Processing on anchors is a result which is preliminary to the problem simplification in the non-anchor case, and is therefore presented first.

\subsection{Anchor nodes, $i\in\C A$}

For \emph{anchor} nodes, due to the constraint $\B x_{i,i}=\B a_i$, problem \e{AL16} becomes separable, that is, it reduces to the parallel of $N_i$ problems of the form
$$
 \B x_{i,j}  \in \argmin_{\Bs x\in\C R} \fract12 (\|\B x-\B a_i\|-r_{i,j})^2 + c_{i} \|\B x-\B y_{i,j}\|^2
\e{AL50}
$$
for $j\in\C N_i$. When dealing with the original non-convex formulation \e{HY8}, a closed-form solution to \e{AL50} can be easily derived from the zero-gradient condition (see the first of \e{AL32})
$$
2c_i (\B x_{i,j}-\B y_{i,j}) = (\B a_{i}-\B x_{i,j}) \left(1-\frac{r_{i,j}}{\|\B a_{i}-\B x_{i,j}\|}\right)\;.
\e{AL52}
$$
By setting $\B x_{i,j} = \B a_i + \alpha \B u$ with $\|\B u\|=1$ and $\alpha>0$, the condition \e{AL52} turns into
$$
\B u (\alpha(1+2c_i)-r_{i,j}) = 2c_i(\B y_{i,j}-\B a_i)\;,
\e{AL54}
$$
providing 
$$
\B u = \frac{\B y_{i,j}-\B a_i}{\|\B y_{i,j}-\B a_i\|}\;,\quad \alpha = \frac{r_{i,j}+2c_i \|\B y_{i,j}-\B a_i\|}{1+2c_i}\;.
\e{AL56}
$$
For the convexified version \e{HY8}, constraint \e{AL52} turns into
$$
2c_i (\B x_{i,j}-\B y_{i,j}) = (\B a_{i}-\B x_{i,j}) \left[1-\frac{r_{i,j}}{\|\B a_{i}-\B x_{i,j}\|}\right]^+\;,
\e{AL52b}
$$
so that the counterpart to \e{AL56} holds with $\alpha$ defined by 
$$
2c_1\alpha + [\alpha-r_{i,j}]^+  = 2c_i \|\B y_{i,j}-\B a_i\| \;,
\e{AL56b}
$$
which ensures
$$
\alpha = \cases{\hbox{\e{AL56}} & if $\|\B y_{i,j}-\B a_i\|>r_{i,j}$\cr
\|\B y_{i,j}-\B a_i\| & otherwise.
}
\e{AL56c}
$$
By putting the above all together, we have
$$
\B x_{i,j} = \cases{\displaystyle \B a_i + \frac{r_{i,j}+2c_i\|\B y_{i,j}-\B a_i\|}{1+2c_i}\frac{\B y_{i,j}-\B a_i}{\|\B y_{i,j}-\B a_i\|}\hspace*{-40mm}\cr
&\rule{0mm}{4mm}if $F_i$ is used, or\cr
& if $\tilde{F}_i$ is used and $\|\B y_{i,j}-\B a_i\|>r_{i,j}$\cr
\rule{0mm}{5mm}\B y_{i,j} & otherwise
}
\e{AL60}
$$
which covers both the convexified case, \e{HY4}, and the non-convex case, \e{HY8}. A graphical interpretation of the result is given in \fig{HY2}.\Fig[h]{HY2}

Observe that, solving the local problem in anchor nodes requires very simple operations. Furthermore, the computational burden carried by \e{AL60} can be transferred from the anchor node to its neighbor nodes. Note in fact that the average action of the product $\B A_i^T$ in the definition of $\B y_i$, in line~18 of \alg{ALO4}, is not needed since $\B y_{i,i}$ is not used. Such a transfer is a reasonable choice whenever the anchor is connected to a large number of nodes, in which case a large overhead in communication is avoided.

\subsection{Nodes which are not anchors, $i\not\in\C A$}

For nodes which are not \emph{anchors}, the result given by \e{AL60} can be exploited to simplify the complexity of the problem from order $n(1+N_i)$ to order $n$. In fact, for a fixed choice of $\B x_{i,i}$ the local solution for $\B x_{i,j}$ can be obtained from \e{AL60} by simply replacing $\B a_i$ with $\B x_{i,i}$. This ensures that
$$
\B x_{i,j}= \B y_{i,j}  + \frac{ \B x_{i,i} -\B y_{i,j}}{\|\B x_{i,i}-\B y_{i,j}\|} \cdot \frac{[\|\B x_{i,i}-\B y_{i,j}\|-r_{i,j}]^\bullet}{1+2c_i} 
\e{AL62}
$$
holds, where we used \e{AL62b}. By substitution in \e{AL16} we obtain an optimization problem in variable $\B x_{i,i}$ only, that is
$$
\B x_{i,i}  =  \argmin_{\Bs x} \fract12\|\B x-\B y_{i,i}\|^2 + \sum_{j\in\C N_i} \frac{ \fract12\big([\|\B x-\tilde{\B y}_{i,j}\|-r_{i,j}]^{\bullet}\big)^2 }{\tilde{c}_{i,j}N_i}
\e{AL64}
$$
where 
$$
\tilde{\B y}_{i,j} = \cases{\B y_{i,j} & $j\not\in\C A$\cr \B a_j & $j\in\C A$}\;,\quad
\tilde{c}_{i,j} = \cases{2c_i & $j\not\in\C A$\cr 1+2c_i & $j\in\C A$}
\e{AL64b}
$$
to separately take into account for the cases where the neighbor $j$ is or is not an anchor. This is a convex problem only for $\tilde{F}_i$. For large $c_i$, however, the function tends to $\|\B x-\B y_{i,i}\|^2$, which is convex by construction in any case. 

Because of the very limited dimension of the problem ($n$ in fact is at most equal to $3$), the local optimization problem given by \e{AL64} is an easy task which can be accomplished by standard optimization methods. In general an algorithm relying on the method of Newton can be chosen to obtain a fast convergence, in which case we will be using the gradient and Hessian 
$$
\eqalign{
\nabla & = \B x-\B y_{i,i} + \sum_{j\in\C N_i} \frac{[\|\B q_j\|-r_{i,j}]^{\bullet}}{\tilde{c}_{i,j}N_i\|\B q_j\|} \B q_j\cr
\nabla^2 & = \B I_n \left(1+  \sum_{j\in\C N_i}  \frac{[\|\B q_j\|-r_{i,j}]^{\bullet}}{\tilde{c}_{i,j}N_i\|\B q_j\|}  \right)\cr
 & \qquad\qquad + \sum_{j\in\C N_i} \frac{r_{i,j} \B q_j\B q_j^T}{\tilde{c}_{i,j}N_i\|\B q_j\|^3}1^\bullet(\|\B q_j\|-r_{i,j})\;,
}
\e{AL66}
$$
where $\B q_j=\B x-\tilde{\B y}_{i,j}$, and where we used \e{AL64b}. Note that, because of the very limited dimension $n$, inversion of the Hessian does not constitute a bottleneck for implementation. Also note that, thanks to the convexity property, the method of Newton leads to an exact result when $\tilde{F}_i$ is used. In the transition to the non-convex function $F_i$, no guarantee is in general available that the global minimum is reached, unless $c_i$ is so large that the problem has become convex (and, incidentally, this latter property guarantees that the distributed algorithm will converge in any case). It is however reasonable to expect that the final outcome will improve over the solution to the convexified problem, as we will discuss in detail in \sect{SI}.

\Section[SI]{Performance Evaluation and Discussion}

Performance of the proposed method is tested on the networks previously used in \cite{Erseghe:sipn15}. Specifically, these are the $N=40$ nodes network with $|\C A|=10$ anchors depicted in \cite[Fig.~2]{Erseghe:sipn15}, and two larger benchmark tests available in Standford's Computational Optimization Laboratory web site \cite{scol}, namely a $N=500$ node network with $|\C A|=10$ anchors, and a $N=1000$ node network with $|\C A|=20$ anchors. All these networks are assumed to have nodes distributed over a $1\times1$ square area, so that the coordinate dimension is $n=2$. For the smaller network of size $N=40$ noisy distance measurements are generated according to an \ac{AWGN} model with standard deviation $\sigma=0.1$ (moderate noise level) and $\sigma=0.01$ (low noise level). For the two larger networks, the noisy distance measurements given by the benchmark tests were used, which correspond to \ac{AWGN} noises with standard deviation of, respectively, $\sigma=0.02$ and $\sigma=0.007$. 

In order to provide a complete insight on the relation of \alg{ALO4} with the solutions already available from the literature, the following algorithms are compared:
\begin{enumerate}
\item SF, namely the \emph{simple and fast} method of \cite{Soares:tsp15} implemented via Nesterov's method;
\item \ac{ADMM}-SF, namely the SF method implemented via the \ac{ADMM} approach of \alg{ALO4} where only the convex relaxation $\tilde{F}_i$ is used, and the transition to the original non-convex formalization is not activated;
\item \ac{SDP}, namely the \ac{SDP} algorithm proposed by \cite{Simonetto:tsp14};
\item \ac{ADMM}-NC, namely the non-convex approach proposed in \cite{Erseghe:sipn15};
\item \ac{ADMM}-H, where H stands for \emph{hybrid}, namely the full implementation of \alg{ALO4}, with an active transition from convex to non-convex functions.
\end{enumerate}

All algorithms are implemented in \emph{MATLAB}, and the minima to local optimization problems \e{AL62}-\e{AL64b} are identified by the \emph{fminunc} solver provided by the \emph{MATLAB Optimization Toolbox}. System performance is evaluated via the \ac{RMSE} measure
$$
{\rm RMSE} = \sqrt{\frac1N \sum_{i\in\C N} \|\B x_{i,i}-\B p_i\|^2}\;,
\e{UI2}
$$
with $\B p_i$ the true position and $\B x_{i,i}$ the estimate locally available at node $i$. System parameters are separately optimized for each algorithm, in order to minimize the number of iterations required for convergence to the optimal point. A detailed prospect on the more relevant chosen parameters is given in \tab{THY2}.\Tab[h]{THY2} To these it must be added that $\rho=\fract15$ was selected for the \ac{SDP} algorithm,  $\lambda_{\max}=10^3$ was selected for all the \ac{ADMM} based algorithms, and $\delta_c=1.01$  and $\theta_c=0.98$ were chosen for \ac{ADMM}-NC and -H. The starting point is set to the all zero vector $\B x_{i,i}=\B 0$, that is, we are investigating a worst case solution where no a priori information is available.

A complete view on \ac{RMSE} performance for the various algorithmic approaches is given in \fig{HY4},\bFig[t]{HY4} where \fig{HY4}.(a) and (b) refer to the smaller network of $N=40$ nodes (two different noise levels), while \fig{HY4}.(c) and (d) refer to the larger networks of, respectively, $N=500$ and $N=1000$ nodes. For the smaller network of $N=40$ nodes the \ac{RMSE} value was obtained as an average performance over $50$ noise realizations, while for the other networks an average operation is already ensured by the fact that we are considering a large number of nodes. Note that the behavior is essentially equivalent in all the considered cases, with the only major distinction that a smaller noise contribution ensures a gain in localization accuracy. The important aspects to be observed are the following. 

We first note that, in the comparison between the SF and the \ac{ADMM}-SF algorithm, the \ac{ADMM} approach provides a significant advantage in terms of convergence speed. This behavior may seem to contradict the widely known fact that, if $t$ is the iteration number, the \ac{ADMM} is known to exhibit $O(1/t)$ convergence speed \cite{goldstein2014fast}, while for Nesterov's method the expected speed is $O(1/t^2)$ \cite{Soares:tsp15}. This is only apparently unreasonable since the convergence rate of Nesterov's method is referring to a convergence to the \emph{exact} minimum point, while for \ac{ADMM} it is capturing the convergence in the dual domain while the convergence to the exact minimum point in the primal domain is known to be exponential \cite{Iutzeler16}. Moreover, it must be observed that in a noisy scenario the \ac{RMSE} performance rapidly saturates in the vicinity of the exact minimum. We can therefore conclude that \ac{ADMM} is better coordinating the local processing exchange to rapidly move towards the vicinity of the minimum point, and in this sense is to be preferred (this effect is known, e.g., see \cite{Boyd:10}). We also note that, since the distributed algorithms we are comparing wirelessly exchange information at each iteration, reducing the total number of iterations implies a lower number of communications between nodes, and, therefore, a limitation in energy consumption. A significant saving is also obtained by \ac{ADMM}-SF in the total time spent for processing, since communication takes a remarkable part of it. The cost to be paid is a more significant local processing effort, which, however, is far below the effort of transmitting and receiving a single packet, which implies a large number of complex operations such as coding, decoding, synchronization, channel estimation, etc. The local computational increase is estimated to be of a factor $N\sub{it}$ equal to the number of  iterations required for the local minimization algorithm \e{AL16} (or, better, of its low complexity counterpart \e{AL64}) to converge. These are expected to be limited and, as a matter of fact, they were set in simulations to be upper bounded by $N\sub{it}\le3$ (further insights on this issue are given later in \fig{HY8}).

The second important aspect given in \fig{HY4} is the performance comparison between the (different) relaxation methods used by the SF and the \ac{SDP} algorithm. It is evident from the figure that \ac{SDP} is more reliable, which confirms the findings of \cite[Fig.~3]{Soares:tsp15}. \ac{SDP} is in any case a very heavy algorithm, and for this reason less suited for implementation (see also \cite{Erseghe:sipn15}). This is also the reason that prevented us from being able to apply  \ac{SDP} to the two larger networks. 

We can finally draw some significant conclusions by observing the relation between \ac{ADMM} based algorithms. Specifically, from \fig{HY4} we clearly see that the proposed hybrid approach given by \ac{ADMM}-H is able to closely follow the rapidly converging behavior of \ac{ADMM}-SF in its initial iterations, and to successively improve the \ac{RMSE} performance by setting itself to the (quasi optimal) performance of \ac{ADMM}-NC. To certify the performance quality, the optimal target given by the \ac{CRLB} (derived according to \cite{Patwari:spm05}) is also reported in \fig{HY4}. With the optimum parameters setting of \fig{HY4} the gain in convergence speed between \ac{ADMM}-H and \ac{ADMM}-NC is limited to a factor of $2$. It is however fair to observe that \ac{ADMM}-H is much more resilient to the choice of parameters. This aspect is investigated in \fig{HY10} which is illustrating that the performance of both \ac{ADMM}-H and \ac{ADMM}-SF are only loosely dependent on the parameters choice, i.e., that suboptimal parameters may simply lead to a (slightly) slower convergence. \ac{ADMM}-NC is instead much more selective, in the sense that parameters choices outside the optimum region may undermine convergence to a good solution or may significantly increase the convergence time. This, in \fig{HY10}, is true in the region $\zeta_c>0.25$.\Fig[t]{HY10}

Some further insights on \ac{ADMM}-H are given in \fig{HY6} and \fig{HY8}. \fig{HY6} illustrates the transition between convex and non-convex formulations by showing the percentage of nodes which use the non-convex functions $F_i$ as a function of the iteration number. Note that the transition is much more sudden for the smaller network of $N=500$ nodes, which in fact requires only $20$ iterations to converge. \fig{HY8} instead illustrates the computational time spent, per iteration, on both \ac{ADMM}-H and SF. Both algorithms were implemented in \emph{MATLAB} in a way to make the time calculation fully comparable. Note that the difference is roughly a factor of $4$ both in the calculation of the \emph{maximum} and the \emph{average} times (with maximum and average taken with respect to times separately calculated on each node). This is in accordance with the fact that $N\sub{it}\le3$. Although not shown in figure we also observe that, as one can expect, the computational times of \ac{ADMM}-NC and \ac{ADMM}-SF (using the efficient formulation \e{AL64}-\e{AL66}) are essentially equivalent to those of  \ac{ADMM}-H.\Fig[t]{HY6}\Fig[t]{HY8}

A final insight is given in \fig{HY14}\Fig[t]{HY14} in a \emph{tracking} context where the network of $N=500$ nodes is moving and the localization algorithm is applied starting from the solution available from the previous step, except made for the initial localization which starts from an all zero vector. Nodes are assumed to be moving on a random direction, with a velocity taken from a Gaussian distribution, and over a square area. If the square area is assumed to have a $100\,$m side, then the average velocity is assumed to be $5\,$km/h (walking speed), the standard deviation is set to $3.33\,$km/h, and the maximum speed to $15\,$km/h. The network is built in such a way that  nodes within $8.33\,$m (or $25\,$m for anchors) are exchanging ranging measurements, and ranging measurements are affected by a standard deviation of $1.66\,$m (equivalent to that of the original test given in \cite{scol}). Nodes are ensured to have access to at least four ranging measurements. Two different algorithms are compared, namely \ac{ADMM}-H and \ac{ADMM}-SF which use the parameters of \tab{THY2}. The ranging and localization measures are updated every second, which corresponds to a step, and each algorithm performs $t=20$ iterations per step. Note from \fig{HY14} that, apart from the very first steps where the algorithms are slowly converging to their respective target performance (this is due to the fact that the maximum number of iterations per step is kept small), then \ac{ADMM}-H sets itself to a \ac{RMSE} performance which is approximately four times lower than that of its \ac{ADMM}-SF counterpart. The optimum level given by the \ac{CRLB} is also shown to certify the estimation quality.

\Section[CO]{Conclusions}

In this paper we built upon the SF convex relaxation method proposed in \cite{Soares:tsp15} and ameliorated it in two ways. First, by casting the problem into a suitable \ac{ADMM} formalization, we were able to identify a fully distributed localization algorithm (\ac{ADMM}-SF) which is scalable, and which sensibly improves in convergence speed over the original proposal based on Nesterov's method. Second, by forcing a transition to the original non-convex function under local convergence, we were able to identify a hybrid algorithm (\ac{ADMM}-H) with improved localization accuracy. Both algorithms were shown to be robust to parameters choices, to be suited for tracking purposes where the number of iterations  per localization step is limited, and to be scalable in that they guarantee fast convergence and accuracy also with large networks. The proposed algorithms were also shown to have limited computational complexity, which guarantees their usability in practical contexts where energy consumption is an issue.


\bibliographystyle{IEEEtran}
\bibliography{hy}

\end{document}